\documentstyle[12pt]{article}

\begin{document}
\setcounter{page}{0}
\newcommand{\al}{\alpha}
\newcommand{\s}{\sigma}
\renewcommand{\L}{\Lambda}
\renewcommand{\b}{\beta}
\renewcommand{\c}{\chi}
\renewcommand{\d}{\delta}
\newcommand{\D}{\Delta}
\newcommand{\wt}{\widetilde}
\renewcommand{\thefootnote}{\fnsymbol{footnote}}
\newcommand{\CL}{{\cal L}}
\newcommand{\pl}[3]{, Phys.\ Lett.\ {{\bf #1}} {(#2)} {#3}}
\newcommand{\np}[3]{, Nucl.\ Phys.\ {{\bf #1}} {(#2)} {#3}}
\newcommand{\pr}[3]{, Phys.\ Rev.\ {{\bf #1}} {(#2)} {#3}}
\newcommand{\prl}[3]{, Phys.\ Rev.\ Lett.\ {{\bf #1}} {(#2)} {#3}}
\newcommand{\ijmp}[3]{, Int.\ J.\ Mod.\ Phys.\ {{\bf #1}} {(#2)} {#3}}
\newcommand{\mpl}[3]{, Mod.\ Phys.\ Lett.\ {{\bf #1}} {(#2)} {#3}}
\newcommand{\zp}[3]{, Z.\ Phys.\ {{\bf #1}} {(#2)} {#3}}
\newcommand{\ap}[3]{, Ann.\ Phys.\ {{\bf #1}} {(#2)} {#3}}
\newcommand{\rmp}[3]{, Rev.\ Mod.\ Phys.\ {{\bf #1}} {(#2)} {#3}}
\newpage
\setcounter{page}{0}
\begin{titlepage}
\begin{flushright}
\hfill{YUMS 95-014}\\
\hfill{SNUTP 95-053}\\
\end{flushright}
\vspace{2.0cm}
\begin{center}
{\Large\bf Large $p_{_T}$ Hadroproduction of $Z$ as a Probe of 
Gluon Distribution inside Proton}\\
\hfill{}
\vskip 0.3cm
{\bf C. S. Kim \footnote{e-mail : kim@cskim.yonsei.ac.kr}
and  Jake Lee \footnote{e-mail : jilee@theory.yonsei.ac.kr}}

{\sl Department of  Physics, Yonsei University,\\
Seoul 120-749, Republic of Korea}
\vskip 0.7cm
\end{center}
\setcounter{footnote}{0}
\begin{abstract}

The transverse momentum distribution of single vector boson
production at hadron colliders provides useful ways
of testing the Standard Model and searching new physics beyond the
Standard Model. We study large $p_{_T}$ hadroproduction of $Z$-boson
as a probe of gluon distributions inside proton.
We investigate how to
get initial gluon-involving contributions, or how to subtract 
quark-quark (or -antiquark) contributions from total cross section. 
We also investigated the simultaneous measurement of the
rapidity  and the transverse momentum 
of the produced $Z$ boson, to obtain momentum fractions of initial partons.
And we extracted  relevant uncertainties involving in experimental 
and theoretical analyses.
This large $p_{_T}$ hadroproduction of $Z$
can be used as constraints on analyses of global parton (gluon and quarks)
distribution functions inside proton.

\end{abstract}

\end{titlepage}

\newpage
\renewcommand{\thefootnote}{\arabic{footnote}}
\baselineskip 24pt plus 2pt minus 2pt
\section{\bf Introduction}

The transverse momentum distribution of single vector boson
production provides useful methods
of testing the Standard Model and searching new physics beyond the
Standard Model. Here we would like to investigate mainly the gluon 
distribution inside proton
using large $p_{_T}$ production of $Z$-boson at hadron colliders.

The recent improvements in the data of deep-inelastic lepton-nucleus
scattering and of the Drell-Yan process have allowed a definitive
determination of the quark distribution functions inside the nucleon.
However, the gluon distribution $G(x, Q^2)$ is not well constrained
by these processes, since it only enters as a second-order effect.
On the other hand, the gluon contributes to the lowest order for
vector boson hadroproduction with large transverse momentum and it has
become normal practice to use fixed-target $pp(\bar p)\rightarrow VX$
data to determine the gluon distribution for $x\sim 2p_{_T}/\sqrt s$
(where $V$ represents a vector boson and $p_{_T}$ is the transverse
momentum of the produced vector boson).
The prompt photon production $p\bar p\rightarrow\gamma X$ has been 
already used on that purpose \cite{harri,auren}.
In principle, collider data obtained for $p\bar p\rightarrow\gamma X$
at the Fermilab energy ($\sqrt s = 1.8$ TeV) with $p_{_T}\sim 10$ GeV
could probe the gluon at $x\sim 10^{-2}$. However, a detailed study of
uncertainties in the theorectical predictions of prompt-photon production
at collider energies has shown that such determination will be
difficult for some reasons \cite{baier,csk}.
One problem is the importance of the bremsstrahlung component at small 
$p_{_T}$, in which the photon is radiated from an outgoing quark 
and so occurs in the debris of a hadronic jet. 
A second ambiguity is related to the choice
of scales in the parton distribution functions and the QCD coupling 
strength $\alpha_s$.
For these uncertainties, massive vector boson $W$ or $Z$
hadroproduction is more useful than $\gamma$.
The bremsstrahlung $W$, $Z$ processes are almost negligible; that is
the outgoing quark (or anti-quark) will very rarely fragment into  $W$ or
$Z$. And the choice of scales can be fixed more acceptably, 
for instance, as $Q^2=M_W ^2$ or $M_Z ^2$.

Previously, as an example, it was pointed out that $W^\pm$ production  
could  be used as measures for gluon and heavy quark distributions 
\cite{csk}, which concentrated on the following subprocesses;
\begin{eqnarray*}
&(i)&\; sg \rightarrow cW\\
&(ii)&\; cg \rightarrow cZ.
\end{eqnarray*}
They proposed ($i$) as a measure of the gluon, and the ratio of ($ii$)
to ($i$) to determine the ratio $c/s$ of quark densities.
However, a $W$-boson possibly decays leptonically into $e\nu$.
And because one cannot detect neutrinos, one could confirm $W$ production
only through detecting an electron and large missing $p_{_T}$, 
which usually results in large systematic uncertainties.
On the other hand, a $Z$-boson decays leptonically to $e^+ e^-$, so
we can confirm $Z$ production with less systematic uncertainties.
Pacing with considerable new data of $Z$
hadroproduction at Tevatron, we would like to investigate gluon densities
of proton by using large $p_{_T}$ hadroproduction of $Z$.
At a glance, it seems possible to use the above subprocess, ($ii$)
$cg\rightarrow cZ$, as a probe of gluon.
But it is not so effective, since this subprocess involves the initial 
charm quark, and the resulting cross section would be very small. 
As one can see more details later, we instead use initial $qg$ 
subprocesses, where $q$ represents any of three light quarks.

In this letter, we present the transverse momentum and
rapidity distributions of $Z$ boson, calculated up to the second order 
in QCD coupling.  When the produced $Z$ has small transverse momentum 
$p_{_T}$, there are large logarithms $log (Q^2 /p_{_T} ^2)$, and 
perturbation theory for $d\s /dp_{_T} ^2$ breaks down there.
And one must either perform a resummation or restrict attention to the
total cross section  by integrating analytically in $p_{_T}$.
These techniques have been carried out to the first order in QCD \cite{mart},
and part of the second-order terms have been analyzed in the
\mbox{$p_{_T}\rightarrow 0$}~limit \cite{davies}.
Therefore, we here restrict ourselves to the region of large $p_{_T}$ 
to avoid these difficulties in small $p_{_T}$.
In the references \cite{arnold,ellis}, the full analytic formulae of
$d\sigma/dp_{_T} ^2 dy$ for large $p_{_T}$ vector boson
production up to the second order in QCD had  been  already presented.
Based on those formulae we study the gluon distribution inside proton 
through large $p_{_T}$ hadroproduction of $Z$-boson. 
By including the full $\alpha\alpha_s ^2$
contributions, we are able to considerably reduce the theoretical errors, 
normally associated with the leading order ($O(\alpha\alpha_s)$) results.
We calculated $K\equiv d\s_2 (Z)/d\s_1 (Z)$ as a function of $p_{_T} (Z)$,
where $d\s_n$ denotes the differential cross section $d\s/dp_{_T} (Z)$
including all QCD subprocesses up to
$O(\al\al_s ^n )$ at $\sqrt s = 1.8$ TeV using  $Q^2 = M_Z ^2$,
and found that $K\simeq 1.3\sim 1.4$ for $p_{_T}(Z) \geq 10 GeV$.
The corrections are significant, but show that the
perturbative expansion is reasonable.

Up to the order of $\alpha\alpha _s ^2$, $p\bar p\rightarrow ZX$ process
includes initial $q\bar q~(qq\; or\; \bar q\bar q)$, $qg~(\bar q g)$
and $gg$ scatterings in partonic level.
Among these subprocesses, only $qg~(\bar q g)$ and $gg$ scatterings are
initial gluon-involving parts.  Therefore, one can analyze the gluon distribution by subtracting initial quark-quark contributions 
from the total subprocesses of $Z$ hadroproduction.
At the Tevatron energies the contribution from initial $gg$ part
is very small, and we can ignore in the numerical analyses.

\section{\bf Detailed investigations of gluon distribution}

As mentioned in Section 1, we would like to use the formulae of
reference \cite{arnold} for our detailed analyses.
The authors of \cite{arnold} presented the hadronic cross sections 
up to the order of 
$\al\al_s ^2$ for $A+B\rightarrow \gamma^\ast + jet(s)$,
\[ {d\s_{_{AB}} \over dp_{_T} ^2 dy} = \sum_{i,j} \int dx_1 dx_2
             f_i ^A (x_1 ,Q^2 )f_j ^B (x_2 , Q^2)
          {sd\tilde{\s} \over dtdu}(\al_s (Q^2), x_1 P_1, x_2 P_2)~, \]
where $A$ and $B$ are the initial hadrons, $P_i$ are their momenta, and
$d\tilde{\s}$ is the factorized partonic cross section in the 
$\overline{MS}$ factorization scheme.

The large $p_{_T}$ hadroproduction of $Z$ up to $\alpha\alpha_s^2$ can be 
calculated by the characteristic five groups of diagrams, as follows
\begin{eqnarray*}
&(i)&\; q\bar q \rightarrow gZ,\\
&(ii)&\; \mbox{Next-to-leading diagrams for } q\bar q \rightarrow gZ,\\
&(iii)&\; q\bar q \rightarrow ggZ,\\
&(iv)&\; q\bar q \rightarrow q\bar qZ,\\
&(v)&\; qq \rightarrow qqZ.
\end{eqnarray*}
Diagrams ($i$) are the first-order ones for $Z$ production through
$q\bar q$ annihilation. Parts of the second-order contribution come
from the interference of these diagrams with the one-loop corrections
of diagrams ($ii$). The rest comes from $2$-$jet$ productions accompanying
the $Z$, such as  the diagrams ($iii$).
The diagrams through initial $qg$-scattering  can be
obtained from ($i$), ($ii$) and ($iii$) by crossing them.
The diagrams through $gg$-scattering can be similarly obtained
from ($iii$) alone.
The diagrams ($iv$) give the remaining parts to $q\bar q$-scattering.
And the diagrams ($v$) are for $qq$-scattering for $Z$ hadroproduction.
Therefore, we have three types of contributions for $Z$ hadroproduction
\begin{eqnarray*}
\bullet\quad QQ &\rightarrow& Z + 1 ~or~ 2 jets,\\
\bullet\quad QG &\rightarrow& Z + 1 ~or~ 2 jets,\\
\bullet\quad GG &\rightarrow& Z + 2 jets,
\end{eqnarray*}
where $Q$ represents initial parton of a quark or anti-quark, 
and $G$ for a gluon.

To further investigate our observations, we numerically calculate 
the transverse momentum $p_{_T}$, and the rapidity $y$
distributions of $Z$ at the Tevatron energies including all QCD subprocesses
up to the order $\al\al_s ^2$ in $p\bar p$ collisions.
We use three input parton distribution functions; MRS(A) \cite{mrs},
GRV(HO) \cite{grv} and CTEQ3 \cite{tung}.
Fig.\ 1 shows $p_{_T}$ distributions using three parton distributions
at $\sqrt s =1.8$ TeV, where we use $Q^2 = M_Z ^2$,
and include three light quarks ($u,\;d,\;s$) as initial quark-partons.
As mentioned earlier, initial $GG$-scattering part is very small, and
we do not show it explicitly here. (Of cause, we include $GG$-part
to calculate total cross section.)
We find that their gluon distributions are almost same at $Q^2 \sim M_Z ^2$,
and in Fig.\ 1 the deviations of distributions
are mainly due to the difference of sea-quark distributions. 
We also note from Fig.\ 1 that it looks very important to subtract the initial
$QQ$-contributions to extract contributions involving gluons only.
However, the $p\bar p \rightarrow Z+jets$ process contains many different
subprocesses, and it might be impossible to distinguish the initial 
$QQ$-contributions from the $QG$-ones {\em experimentally} by analyzing
shapes of final hadronic jets, and {\it etc.}
Now that we have theoretical predictions of the initial $QQ$-contributions,
we can in principle subtract those from the experimental total result.
And by comparing the remaining experimental cross section and theoretical
one, one can probe the gluon distributions inside proton.
However, as can be seen in Fig.\ 1 for $p_{_T} \geq 10$ GeV,
$QQ$-contributions are about $70\%$ of total cross section,
and it looks inappropriate to subtract $QQ$-part, which is not a minor part,
but a major contribution. Therefore, one should raise
$p_{_T}$ minimum cut up to about $40\sim 50$ GeV, to make 
$QQ$- and $QG$-contributions at least experimentally comparable. 
These practical considerations will be presented in more detail in Section 3.

Assuming that we could experimentally distinguish $Z+1jet$ from $Z+2jets$, 
let us first consider simple $Z+1jet$ case.
In partonic level, $p\bar p \rightarrow Z+1jet$ process includes 
only two types of diagrams;
\begin{eqnarray*}
\bullet\quad q\bar q &\rightarrow& Zg,\\
\bullet\quad qg &\rightarrow& Zq~({or}~ \bar q g\rightarrow Z\bar q).
\end{eqnarray*}
In these processes, we can extract initial gluon-involving contributions
by subtracting {\em theoretical} prediction of $q\bar q \rightarrow Zg$ 
subprocess from the {\em experimental} total results of $p\bar p \rightarrow Z+1jet$. 
Futhermore, if we could distinguish the {\em quark-jet} from
the {\em gluon-jet} through theoretical and experimental combined analyses,
only the initial gluon contributing part can be extracted experimentally 
from {\em Z + 1 quark-jet}.
Fig.\ 2 shows the $p_{_T}$ distributions of $Z+1jet$ from
$q\bar q\rightarrow Zg$ and $qg\rightarrow Zq$ 
($\bar q g\rightarrow Z\bar q$).
For comparison, we also show $p_{_T}$ distributions of
$p\bar p \rightarrow Z + 1jet$ and $Z + 2jets$.

Until now we have only considered the transverse momentum $p_{_T}$
distributions. We now consider analyses with simultaneous 
measurement of both rapidity $y$ and $p_{_T}$ of the produced $Z$ boson.
The $y$ and $p_{_T}$ of the $Z$ boson in the laboratory 
frame are related to the momentum fractions of initial partons by
\begin{eqnarray}
y &=& \frac{1}{2}\ln\frac{E + p_{_L}}{E - p_{_L}}\nonumber\\
  &=& \ln \left \{ \frac{M_Z ^2 + x_1 x_2 s - s_2}{2x_2 \sqrt s (M_Z ^2
      + p_{_T} ^2)^{1/2}} \pm
      \sqrt{\frac{(M_Z ^2 + x_1 x_2 s - s_2)^2}{(2x_2 \sqrt s (M_Z ^2
      + p_{_T} ^2)^{1/2})^2} - \frac{x_1}{x_2}} \right \},
\end{eqnarray}
where $s$ is the invariant mass of the incoming hadrons, and
$s_2$ is the invariant mass of final two-jets 
{\sl when two jets accompany the produced $Z$}.
And $x_1,\, x_2$ are the four-momentum fractions of the colliding 
partons. If we consider only $Z+1jet$ processes,
the value of $s_2$ in the above relation becomes zero.
By using the above relation we can directly obtain the gluon 
momentum fraction inside incoming proton.

\section{\bf Discussions}

We now study relevant uncertainties in analyzing large $p_{_T}$
hadroproduction of $Z$, which one must
take into account in theoretical and experimental analyses. 
The first uncertainty comes from QCD-scale dependences.
Although there is much less ambiguity associated with the choice of scales
compared to direct photon case, as explained before, there still remains  
weak QCD-scale dependence because of considering only finite order of
perturbative corrections.
\begin{table}[htb]
\centering
\caption{\em The integrated total cross sections with the
same scales for $\alpha_s$ and parton distributions.}
\begin{tabular}{c|cccc}
\hline\multicolumn{5}{c}{$\sigma \, (nb)$ \qquad $p_{_T} >10 \, GeV$,
                   \quad $ |y| < 2.5 $}\\ \hline\hline
    scale $Q^2$ & $M_Z ^2$ & $p_{_T} ^2 /2$ & $p_{_T} ^2$ & $2p_{_T} ^2$\\ \hline
        MRS(A) &  1.84 & 2.16 & 2.06 & 1.99\\
        GRV(HO) &  1.73 & 2.05 & 1.95 & 1.88\\
        CTEQ3 &  1.87 & 2.19 & 2.10 & 2.02\\ \hline
\end{tabular}
\end{table}
In  Table 1 we give total cross sections for $p\bar p \rightarrow ZX$
production integrated over the region $p_{_T}(Z)>p_{min} ^Z=10$ GeV 
for various $Q^2$.
The value of $p_{min} ^Z$ is chosen so as to retain as many clear events
as possible and yet to remain in a region where perturbation theory
gives a reliable prediction. The same scale is chosen for the parton
densities and QCD strength $\al_s (Q^2)$. But these scales need not be 
the same.  Table 2 shows the sensitivity, when we choose different scales 
for $\al_s (Q^2)$ and parton distribution functions. 
We note that there show significant differences
in results between $Q^2=M_Z ^2$ and $Q^2=p_{_T} ^2 /2$ 
for parton distributions, where we fix the scale of 
$\alpha_s (Q^2)$ as $M_Z ^2$, as in  Table 2.
Therefore, once we have sufficient experimental data for this
process ($p\bar p\rightarrow ZX$), we could even investigate scale 
dependence of QCD-coupling $\al_s$, as well as parton distribution functions.
\begin{table}[htb]
\centering
\caption{\em The integrated total cross sections with
different scales for $\alpha_s$ and parton distributions.
   We fix $Q^2=M_Z ^2$ for $\alpha_s$ and use four different scales
   for parton distribution functions.}
\begin{tabular}{c|cccc}
\hline\multicolumn{5}{c}{$\sigma \, (nb)$ \qquad $p_{_T} >10 \, GeV$,
                     \quad $ |y| < 2.5 $}\\ \hline\hline
   scale $Q^2$ & $M_Z ^2$ & $p_{_T} ^2 /2$ & $p_{_T} ^2$ & $2p_{_T} ^2$\\
 \hline
        MRS(A) &  1.84 & 1.52 & 1.59 & 1.64\\
        GRV(HO) &  1.73 & 1.43 & 1.50 & 1.54\\
        CTEQ3 &  1.87 & 1.54 & 1.61 & 1.66\\ \hline
\end{tabular}
\end{table}

Next we still have uncertainties related to structure 
functions of initial quarks, when we try to derive informations on gluon. 
Here we use three different parametrizations; MRS(A), GRV(HO), CTEQ3.
Martin {\it et. al.} \cite{mrs} improved their parametrization of
parton densities through a new global analysis on deep inelastic 
scatterings and related data including  the recent measurements
of $F_2$ at HERA, on the asymmetry of the rapidity distributions of
$W^\pm$ production \cite{elb} at Tevatron, and on the asymmetry
in Drell-Yan production in $pp$ and $pn$ collisions \cite{stir}.
We use their latest version of MRS(A) functions.
CTEQ Collaboration \cite{tung} also improved their distribution functions, incorporating several new types of data and presented the new 
version CTEQ3, which we use here.
Gl\"uck {\it et.\ al.} \cite{grv} predicted the parton
distributions down to $x\simeq 10^{-4}$ and $Q^2 \simeq 0.3$ GeV$^2$,
using the data from deep inelastic scattering experiments at
$x\ge 10^{-2}$ together with the idea that at some low resolution scale
the nucleon consists entirely of valence quarks and {\em valence\/}-like
gluons. They presented parton distributions obtained for the leading
order (GRVLO) as well as for the higher order  (GRVHO)
calculation. We use GRVHO.

The simultaneous measurement of $y$ and $p_{_T}$
would enable us to determine the momentum fractions of initial 
partons from Eq.\ (1), and it was previously  pointed out that 
one should be very careful in dealing with $s_2$,
where $s_2$ is the invariant mass of final state two jets,
\[ s_2=(p_{jet1} + p_{jet2})^2.\]  
If only one jet presents in the final states, 
$s_2=0$ in our massless parton level approximation, as explained before. 
Experimenalists should find here the best optimized 
solution through realistic Monte Carlo studies
to differentiate one-jet from two-jets final states, which are 
accompanying the produced $Z$-boson. 
Fig.\ 3 shows the $p_{_T}$ distributions of
$p\bar p\rightarrow Z+jets$ for $y=0$.

We next calculate numerically the number of events at Tevatron energies. 
For $|y|<2.5$ and $p_{_T} \geq 10$ GeV, we get
\begin{eqnarray*}
 Number\; of\; events / 10 pb^{-1} 
	 &=& [2.00\pm 0.15]\times 10^4 \quad \mbox{for~ MRS(A)},\\
         &=& [1.89\pm 0.16]\times 10^4 \quad \mbox{for~ GRV(HO)},\\
         &=& [2.03\pm 0.16]\times 10^4 \quad \mbox{for~ CTEQ3},
\end{eqnarray*}
where the errors are such that the lower and upper limits correspond 
to the scale choice of $Q^2 = M_Z ^2$ and $Q^2 = p_{_T} ^2 /2$, respectively.
As previously mentioned, about $70\%$ of the above events is from the initial
$QQ$-contributions, and $QG$-contributions, which we are interested in 
for our investigation of gluon distributions, are only $30\%$ of 
total events. It would not be so effective to subtract from total number of events $QQ$-contributions, which are much larger than $QG$-ones.
Therefore, we need raise $p_{_T}$ minimum cut to make initial 
$QG$-contributions at least comparable to $QQ$-ones.
Fig.\ 4 shows the rapidity  distributions with (a) $p_{_T} \geq 10$ GeV,
and (b) $p_{_T} \geq 40$ GeV. 
And for $|y|<2.5$ and $p_{_T} \geq 40$ GeV, we get
\begin{eqnarray*}
 Number\; of\; events / 10 pb^{-1}
	   &=& [2.27\pm 0.07]\times 10^3 \quad \mbox{for~ MRS(A)}, \\
           &=& [2.18\pm 0.07]\times 10^3 \quad \mbox{for~ GRV(HO)}, \\
           &=& [2.35\pm 0.08]\times 10^3 \quad \mbox{for~ CTEQ3}.
\end{eqnarray*}
As we can see in Fig.\ 4(b), about a half of the total number of events
is from initial $QG$-contributions.
Therefore, we can summarize, assuming luminosity $\CL = 10pb^{-1}/year$
\begin{eqnarray*}
 Number\; of\; events/year \;(p_{_T}\geq 10GeV) &=& [1.96\pm 0.08\pm 0.16]
                          \times 10^4, \\
 Number\; of\; events/year \;(p_{_T}\geq 40GeV) &=& [2.27\pm 0.08\pm 0.08]
                          \times 10^3,
\end{eqnarray*}
where the first error is due to structure functions,
and the second is from scale ($Q^2$) dependence.
For $p_{_T}\geq 10$ GeV (see Fig.\ 4(a)), the total errors are about
$9\%$, and the $QG$-contributions are only about $30\%$. 
One the other hand, we note that for $p_{_T}\geq 40$ GeV
the initial $QG$-contributions are comparable to the $QQ$-ones \
(see Fig.\ 4(b)), and the theoretical errors are also reduced to 5\%.
Therefore, for large $p_{_T}$ minimum cut we can investigate gluon
structure functions with less theoretical uncertainties.
Moreover, we can see  from Fig.\ 4(b) that $QG$-contributions
are even larger than $QQ$-ones at large rapidity region. It means that
($Q(x_q)+G(x_g)$)-scattering contributions are dominant at large 
($x_q -x_g$), where $x_q$ and $x_g$ are momentum fractions of intital 
scattering  quark and gluon, with the kinematic constraint of 
$s x_q x_g > M_Z^2$.

Fig.\ 5 shows rapidity distributions at
$\sqrt s = 14$ TeV (LHC energies). At these energies, 
$QG$-contributions exceed $80\%$ of total cross section.
If we consider only $Z+1jet$ case at these LHC energies,
because initial $QG$-scatterings produce only {\em quark-jet} final states,
we can investigate quark fragmentation functions using this
large $p_{_T}$ hadroproduction of $Z$-boson.
For $p_{_T}\geq 50$ GeV, total number of events at ${\sqrt s} = 14$ TeV is 
\[ Number\; of \; events/100 pb^{-1} \; 
=\; [3.86\pm 0.06 \pm 0.11]\times 10^5, \]
where the first (second) errors are due to structure function (QCD-scale)
dependence.

This process has been already used for other purposes.
In  recent papers \cite{cdf}, it is reported that the cross sections
for $W$ and $Z$ production in $p\bar p$ collisions at $\sqrt s =1.8$ TeV
are measured 
at the Fermilab Tevatron colliders for final states; 
$W\rightarrow e\nu_e$, $Z\rightarrow e^+ e^-$,
$W\rightarrow \mu\nu_\mu$, and $Z\rightarrow \mu^+ \mu^-$.
And it is pointed out that assuming the Standard Model couplings,
this result can be used to determine the width and mass of the $W$ boson.

\vskip 2.0cm
\centerline{\bf Acknowledgements}
\medskip

We would like to thank Sun Kee Kim for his careful reading of 
manuscript and for his valuable suggestions in experimental aspects.
And we also thank E. Reya,  W.\ K.\ Tung, A.\ D.\ Martin, R.\ C.\ Roberts
and W.\ J.\ Stirling for kindly sending us their newest parton
structure functions.
The work was supported
in part by the Korean Science and Engineering  Foundation, 
Project No. 951-0207-008-2,
in part by Non-Directed-Research-Fund, Korea Research Foundation 1993, 
in part by the CTP, Seoul National University,
in part by Yonsei University Faculty Research Grant 1995,  and
in part by the Basic Science Research Institute Program, Ministry of 
Education, 1995,  Project No. BSRI-95-2425.

\vfill\eject

\vfill\eject
\centerline{\bf Figure Captions}
\medskip
\begin{description}

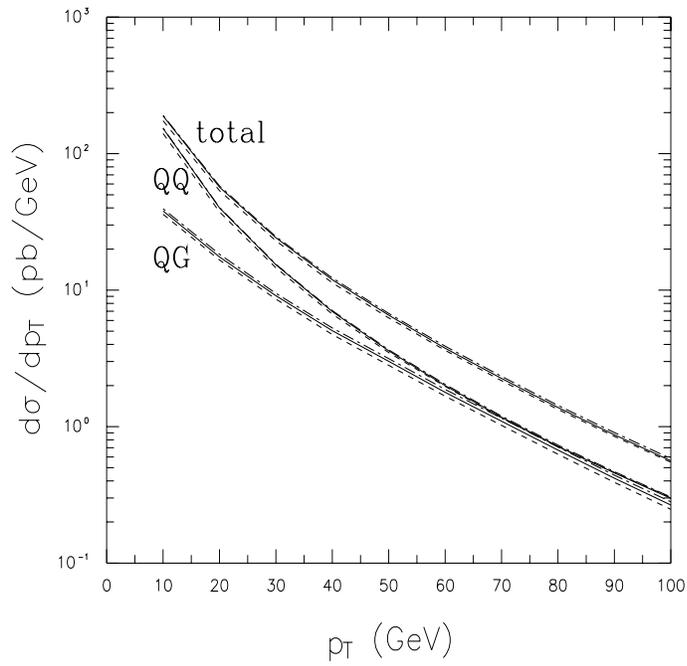
\item[Fig.1] The $p_{_T}$ distributions of $p\bar p\rightarrow Z+X$
     at $\sqrt s = 1.8$ TeV using $Q^2 = M_Z ^2$. $QQ$ and $QG$
     represent the initial $quark-quark$ (or $-antiquark$) 
     and $quark-gluon$ contributions respectively, 
     and $total$ represents for the sum of all contributions.
     The solid line is for MRS(A), the dash-dotted is for CTEQ3,
     and the dashed is for GRV(HO).

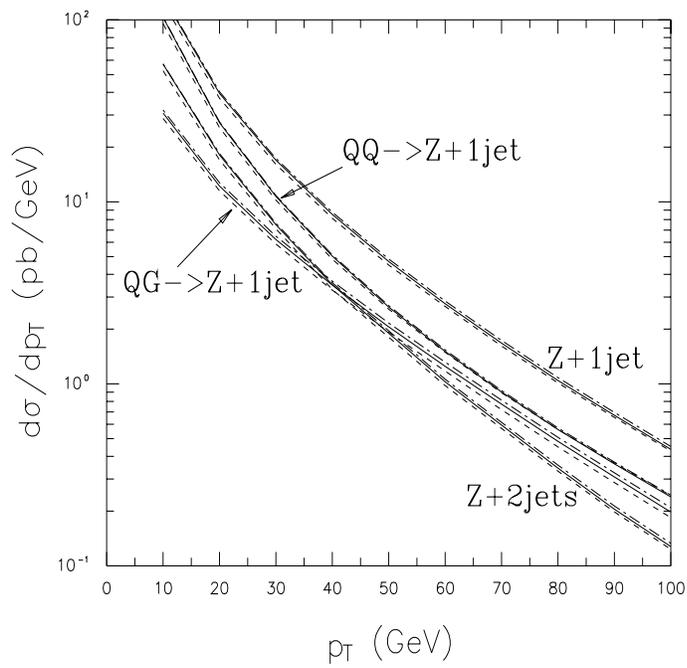
\item[Fig.2] The $p_{_T}$  distributions of $p\bar p\rightarrow Z+1jet$
     from ($QQ$) $q\bar q\rightarrow Zg$ and ($QG$) 
     $qg\rightarrow Zg~(\bar q g\rightarrow Z\bar q)$
     at $\sqrt s = 1.8$ TeV using $Q^2 = M_Z ^2$.
     Also shown are the $p_{_T}$  distributions of
     $p\bar p\rightarrow Z+1jet$ 
     and $Z+2jets$.

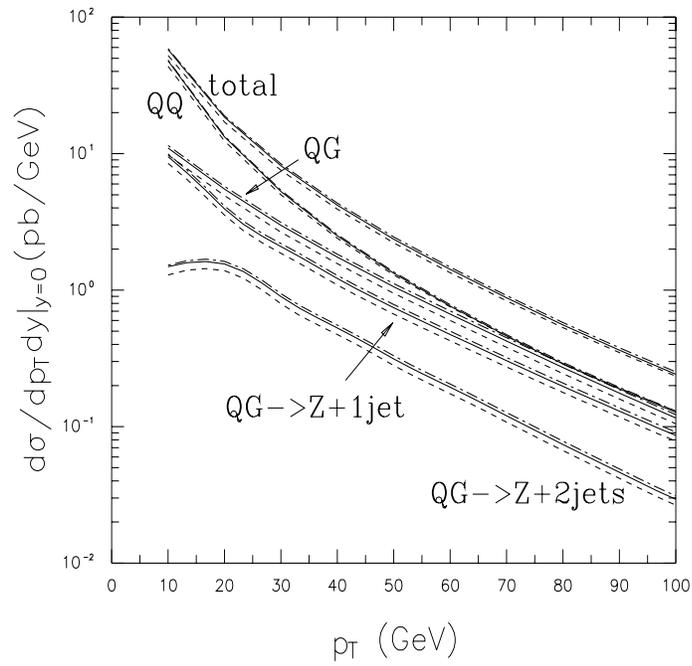
\item[Fig.3] The $p_{_T}$  distributions of $p\bar p\rightarrow Z+X$
     for $y=0$, at $\sqrt s = 1.8$ TeV using $Q^2 = M_Z ^2$.
     Also shown are the $p_{_T}$ distributions of
     $QG\rightarrow Z + 1jet$ and $Z + 2jets$ cases for $y=0$.
    
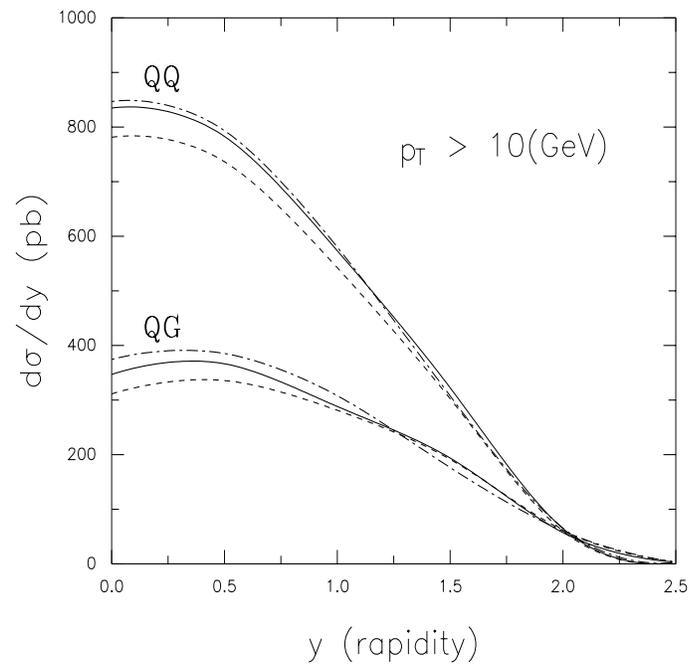
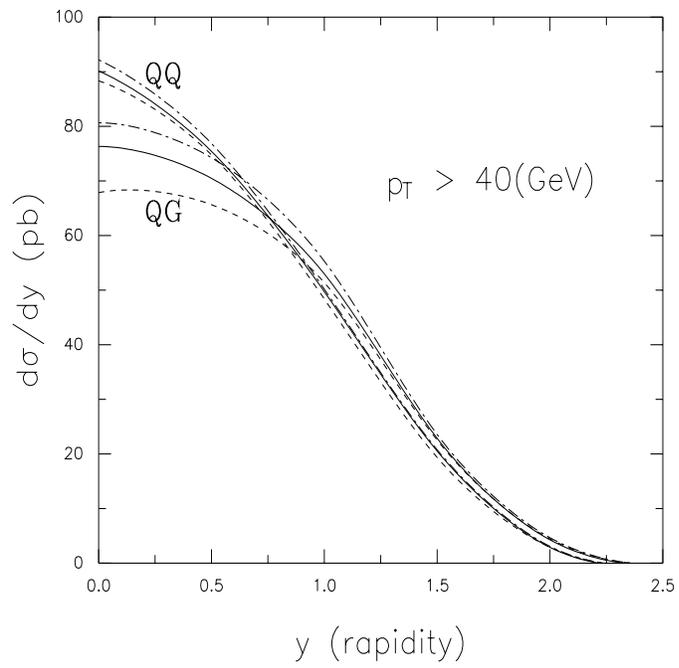
\item[Fig.4] The rapidity distributions of $QQ\rightarrow Z+X$ and
     $QG\rightarrow Z+X$ (a) for $p_{_T}\geq 10$ GeV, 
     and (b) for $p_{_T}\geq 40$ GeV, 
     at $\sqrt s = 1.8$ TeV using $Q^2 = M_Z ^2$.

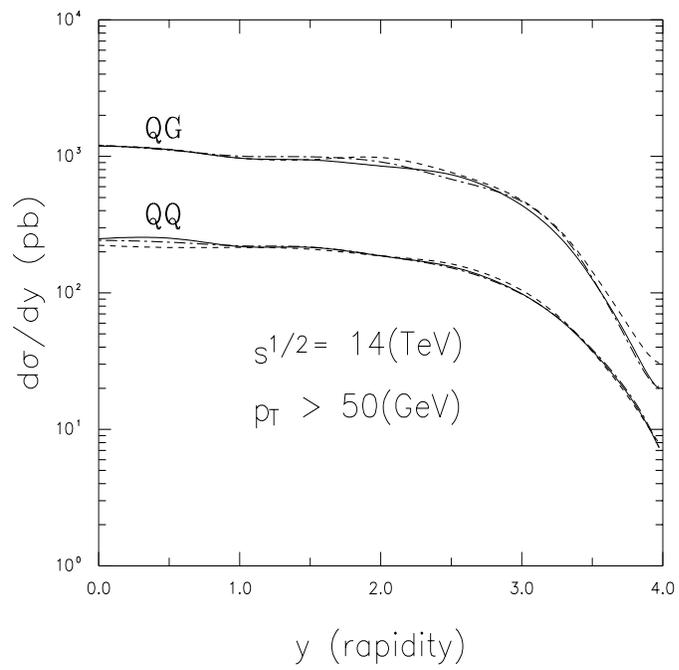
\item[Fig.5] The rapidity distributions of $QQ\rightarrow Z+X$ and
     $QG\rightarrow Z+X$ for $p_{_T}\geq 50$ GeV
     at $\sqrt s = 14$ TeV using $Q^2 = M_Z ^2$.

\end{description}

\end{document}